\documentclass[aip,amsmath,amssymb,floatfix,reprint,citeautoscript]{revtex4-2}
\usepackage[utf8]{inputenc}
\usepackage[T1]{fontenc}
\usepackage{mathptmx}
\usepackage{etoolbox}

\usepackage{graphicx} % figures
\usepackage{bm} % bold math
\usepackage{wrapfig} % wrap text around figure
\usepackage[colorlinks,allcolors=black,citecolor=blue,urlcolor=blue]{hyperref} % appearance of the hyperrefs
\usepackage{booktabs} % nicer tables
\usepackage{silence}
\usepackage{multirow}
\makeatletter
\def\@email#1#2{%
 \endgroup
 \patchcmd{\titleblock@produce}
  {\frontmatter@RRAPformat}
  {\frontmatter@RRAPformat{\produce@RRAP{*#1\href{mailto:#2}{#2}}}\frontmatter@RRAPformat}
  {}{}
}%
\makeatother

\begin{document}

% manuscipt title
\def\mstitle{Non-Aqueous Ion Pairing Exemplifies the Case for Including Electronic Polarization in Molecular Dynamics Simulations}

\title{\mstitle}

\author{Vojtech Kostal}
\affiliation{
Institute of Organic Chemistry and Biochemistry of the Czech Academy of Sciences, Flemingovo nám. 2, 166 10 Prague 6, Czech Republic
}

\author{Pavel Jungwirth*}
\email{pavel.jungwirth@uochb.cas.cz}
\affiliation{
Institute of Organic Chemistry and Biochemistry of the Czech Academy of Sciences, Flemingovo nám. 2, 166 10 Prague 6, Czech Republic
}

\author{Hector Martinez-Seara*}
\email{hseara@gmail.com}
\affiliation{
Institute of Organic Chemistry and Biochemistry of the Czech Academy of Sciences, Flemingovo nám. 2, 166 10 Prague 6, Czech Republic
}

\date{\today}

\begin{abstract}

\setlength\intextsep{0pt}
\begin{wrapfigure}{r}{0.4\textwidth}
  \hspace{-1.8cm}
  \includegraphics[width=0.4\textwidth]{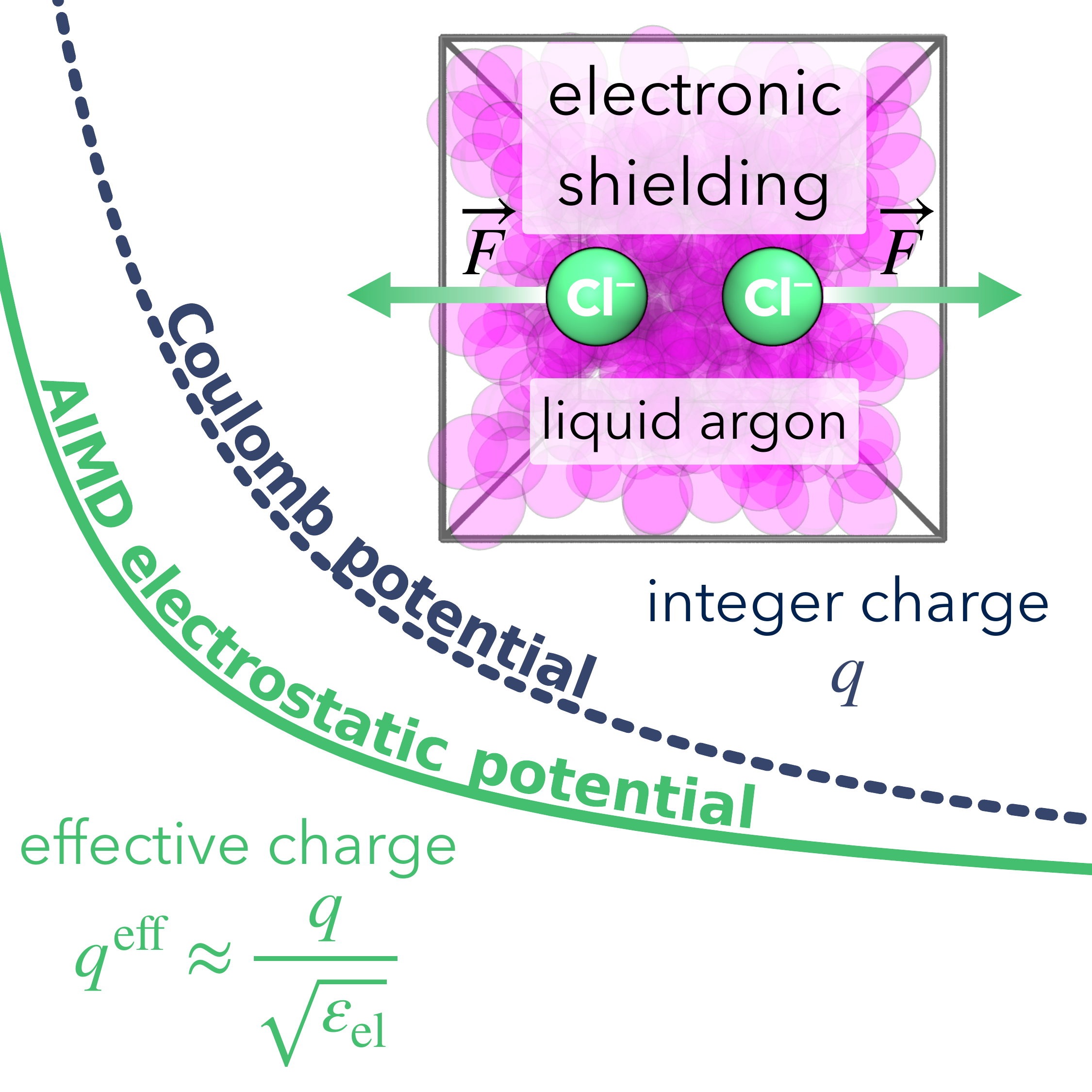}
\end{wrapfigure}

The inclusion of electronic polarization is of crucial importance in molecular simulations of systems containing charged moieties.
When neglected, as often done in force field simulations, charge--charge interactions in solution may become severely overestimated leading to unrealistically strong bindings of ions to biomolecules.
The electronic continuum correction introduces electronic polarization in a mean-field way via scaling of charges by the reciprocal of the square root of the high-frequency dielectric constant of the solvent environment.
Here, we use \textit{ab initio} molecular dynamics simulations to quantify the effect of electronic polarization on pairs of like-charged ions in a model non-aqueous environment where electronic polarization is the only dielectric response.
Our findings confirm the conceptual validity of the present approach, underlining its applicability to complex aqueous biomolecular systems.
Simultaneously, the present results justify the potential employment of weaker charge scaling factors in force field development.

\end{abstract}

\maketitle

The electronic continuum correction (ECC) allows for incorporating electronic polarization effects in a mean-field way into molecular dynamics simulations governed by simple nominally non-polarizable empirical potentials~\cite{Leontyev2009-10.1063/1.3060164,Kirby2019-10.1021/ACS.JPCLETT.9B02652}.
What makes ECC particularly appealing is its ability to accomplish this in an effective and computationally efficient manner without necessitating any alterations to the existing simulation software, being thus truly a ``free lunch`` approach~\cite{Leontyev2011-10.1039/C0CP01971B, Kirby2019-10.1021/ACS.JPCLETT.9B02652, Zeron2019-10.1063/1.5121392, Duboue-Dijon2020-10.1063/5.0017775}.
Empirical non-polarizable potentials capture well nuclear polarization ($\varepsilon_\mathrm{nuc}$).
However, recovering the electronic polarization ($\varepsilon_\mathrm{el}$) is more cumbersome as it rises from electron density changes, which are not accounted for in these force-field models.

Neglecting the effects of the electronic polarization leads to exaggerated electrostatic interactions resulting in quantitative and even qualitative disagreement with experiments in properties such as the strength of ion pairing~\cite{Kirby2019-10.1021/ACS.JPCLETT.9B02652, Duboue-Dijon2020-10.1063/5.0017775}.
While a thorough choice of model parameters can partially address this issue for molecular species, a more comprehensive treatment is essential to account for electronic polarization for ions and charged groups.
One way to remedy the problem in force-field MD is to include electronic polarization explicitly~\cite{Liu2019-10.1021/acs.jctc.9b00261}.
However, this is still not a common practice in biomolecular simulations due to issues connected with parameterization and computational efficiency~\cite{Duboue-Dijon2020-10.1063/5.0017775}.
In contrast, the ECC approach is simple and computationally straightforward.
It involves scaling the integer charges of ions or charged molecular groups by the reciprocal square root of the electronic permittivity ($\varepsilon_\mathrm{el}$, \textit{i.e.}, the high-frequency dielectric constant) of the surrounding environment.
This accounts for electronic polarization in a mean-field way~\cite{Leontyev2009-10.1063/1.3060164, Leontyev2010-10.1021/CT9005807, Leontyev2011-10.1039/C0CP01971B} by immersing the whole system in a dielectric continuum with the high frequency dielectric constant $\varepsilon_\mathrm{el}$.

The charge scaling relation then emerges directly from the Coulomb potential (where $q_1$ and $q_2$ are the two integer charges separated by a distance $r_{12}$ and $\varepsilon_0$ is the permittivity of vacuum)
\begin{equation}
    \label{eq:scaled-coulomb}
    U_\mathrm{C} = \frac{1}{4\pi \varepsilon_0}\frac{q_1q_2}{\varepsilon_\mathrm{el}}\frac{1}{r_{12}}
      = \frac{q^\mathrm{eff}_1q^\mathrm{eff}_2}{4\pi \varepsilon_0 r_{12}},
\end{equation}
\begin{equation}
    \label{eq:scaled-charge}
    q^\mathrm{eff}_\mathrm{1,2} = \frac{q_\mathrm{1,2}}{\sqrt{\varepsilon_\mathrm{el}}}.
\end{equation}

The applicability of the ECC method is based on the assumption of electronic homogeneity of the system.
While biological systems are typically strongly non-homogeneous in terms of the total dielectric constant, the high-frequency component is almost constant.
Indeed, for both aqueous and non-polar biological environments, $\varepsilon_\mathrm{el}$ varies from about 1.7 to 2.2 (see Table~I in Reference~\citenum{Duboue-Dijon2020-10.1063/5.0017775}), which corresponds to scaling factor ranging from 0.77 to 0.67 (with that for water equal to 0.75).

Charge scaling was shown to improve the description of a wide range of systems including ionic liquids~\cite{Cui2019-10.1021/acs.jpcb.9b08033}, aqueous ionic solutions~\cite{Blazquez2023-10.1063/5.0136498}, aqueous biomolecular systems interacting with ions~\cite{Duboue-Dijon2020-10.1063/5.0017775, Melcr2019-10.3389/fmolb.2019.00143}, ions at polar/nonpolar interfaces~\cite{Vazdar2012-10.1021/JZ300805B}, ions adsorbed to metal--oxide surfaces~\cite{Biriukov2018-10.1039/C8CP04535F, Biriukov2019-10.1021/acs.langmuir.8b03984, Biriukov2020-10.1021/acs.jpcc.9b11371}, and osmotic and activity coefficients~\cite{Bruce2018-10.1063/1.5017101} in systems where charge--charge interactions are important.
Although ECC has a firm physical foundation, the employed scaled charges have not been directly validated yet, and the framework itself is still subject to debate~\cite{Chaumont2020-10.1021/ACS.JPCB.0C04907}.
In several recent studies, the scaling factor has been treated as an adjustable parameter rather than being directly derived from the value of $\varepsilon_\mathrm{el}$~\cite{Kann2014-10.1063/1.4894500, Zeron2019-10.1063/1.5121392}.
To quantify the electronic screening of ionic charges in solutions, \textit{ab initio} molecular dynamics (AIMD) is the tool of choice due to its explicit evaluation of electronic polarization.
In this context, an earlier study of ionic liquids~\cite{Zhang2012-10.1021/JP3037999}, where the fitting of the AIMD electrostatic potential in the liquid phase yielded scaled charges, already lent indirect support to the ECC concept~\cite{Zhang2012-10.1021/JP3037999}.

In this work, we employ AIMD simulations quantifying the degree of attenuation of charge--charge interaction between ions due to the electronic permittivity of the solvent environment.
By calculating the free energy profiles of ion pairing and extracting the contribution due to electronic polarization, we obtain in an unbiased way the charge scaling factor as a function of the interionic separation.
In this way, we provide a solid foundation for further development of the ECC framework, also assessing the robustness of the mean-field approximation employed within ECC.

To reach the above goals, we quantify ion pairing in an environment that exhibits electronic polarization as the only dielectric response to the presence of ions, namely in liquid argon.
In previous studies, ECC successfully reproduced free energy profiles of ion pairing obtained by the AIMD~\cite{Pluharova2013-10.1021/JZ402177Q, Martinek2018-10.1063/1.5006779}, but the effect of electronic polarization could hardly be rigorously separated from other electrostatic contributions.
While chemically distant from water, liquid argon (as many other liquids) possesses electronic permittivity comparable to water.
Therefore, the choice of the present relatively simple system is relevant for quantifying charge scaling and electronic polarization effects in solvents in general.
According to the experimental refractive indices~\cite{CRCHandbook-2016-refraction-index, Sinnock1969-10.1103/PhysRev.181.1297}, liquid water and argon have the high-frequency dielectric constants of mutually close values of $\varepsilon_\mathrm{el}=1.78$ and $\varepsilon_\mathrm{el}=1.52$, corresponding to similar scaling factors of 0.75 and 0.81.
Thanks to the fact that argon (unlike water) lacks a permanent dipole or higher electrostatic moment,
the static (nuclear) dielectric constant $\varepsilon_\mathrm{nuc}$ equals to one.
This grossly simplifies our objective of rigorously extracting charge scaling factors as a function of interionic distance as only $\varepsilon_\mathrm{el}$ attenuates the electrostatic interactions.

Studying ions in liquid argon instead of water enhances the convergence as it is a simple Lennard-Jones liquid.
Still, it brings unique challenges despite being a seemingly trivial system.
Argon is a poorly stabilizing medium for ions of opposite charges, and one can hardly avoid spurious charge transfer from anion toward cation when using electronic structure methods such as the density functional theory (DFT).
To avoid charge transfer between ions of opposite charge (which would obscure extraction of charge scaling factors), we employ AIMD to systems comprising of a like charge pair of ions in liquid argon.
When simulating two ions with like charges, it is necessary to neutralize the net charge by adding a uniform background charge of opposite sign when accounting for the long-range electrostatics.
In the SI, we demonstrate that for interionic separations and system sizes studied here, our results are not significantly affected by the effect of the neutralizing background charge.

\begin{figure}[b]
    \centering
    \includegraphics[width=\linewidth]{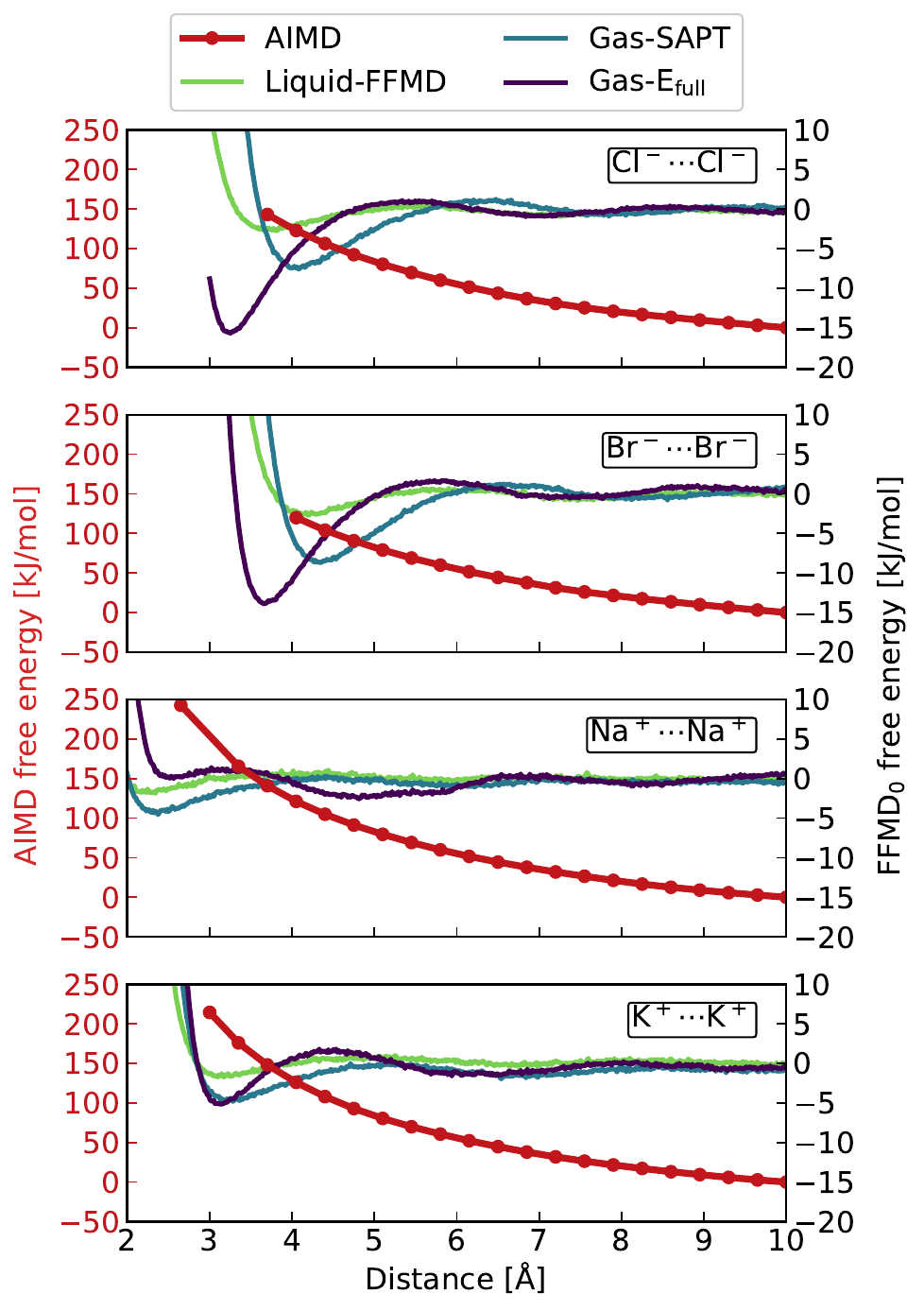}
    \caption
    {
    Free energy as a function of the ion--ion separation obtained by AIMD (red, left $y$-axis) and FFMD (right $y$-axis).
    The latter uses zeroed ionic charges employing non-bonded parameters derived in various manners described in Computational Details: Liquid-FFMD (green), Gas-SAPT (blue), Gas-E$_\mathrm{full}$ (purple).
    Error calculated by bootstrapping of the AIMD free energy amounts to 10.5, 8.9, 8.6, 8.7~kJ/mol for chloride, bromide, sodium, and potassium, respectively.
    }
    \label{fig:free-energies}
\end{figure}

We have obtained the four free energy profiles of ion pairing for two sodium, potassium, chloride, or bromide ions in liquid argon using AIMD by integrating the mean force along a set of distances ranging from the close ion--ion contact up to a separation of 10~\AA.
The AIMD free energy profiles are shown as red lines in Figure~\ref{fig:free-energies}.
To remove the van der Waals and entropy contributions to the \textit{ab initio} free energy we subtracted from the AIMD curves auxiliary force-field molecular dynamics (FFMD) free energy profiles with zeroed ionic charges.
This subtraction scheme relies on setting the van der Waals interactions in the FFMD simulation to mimic the AIMD counterpart, ensuring that the two free energy profiles have comparable non-electrostatic contributions.
To check the robustness of this procedure (since there is no unique way how to do this mapping), the van der Waals interactions were modeled by Lennard-Jones (LJ) (12--6) potentials using parameters obtained by three distinct approaches denoted here as ``Liquid-FFMD``, ``Gas-SAPT``, and ``Gas-E$_\mathrm{full}$`` 
(further description is provided in the Computational Details section).

\begin{figure}[b]
    \centering
    \includegraphics[width=\linewidth]{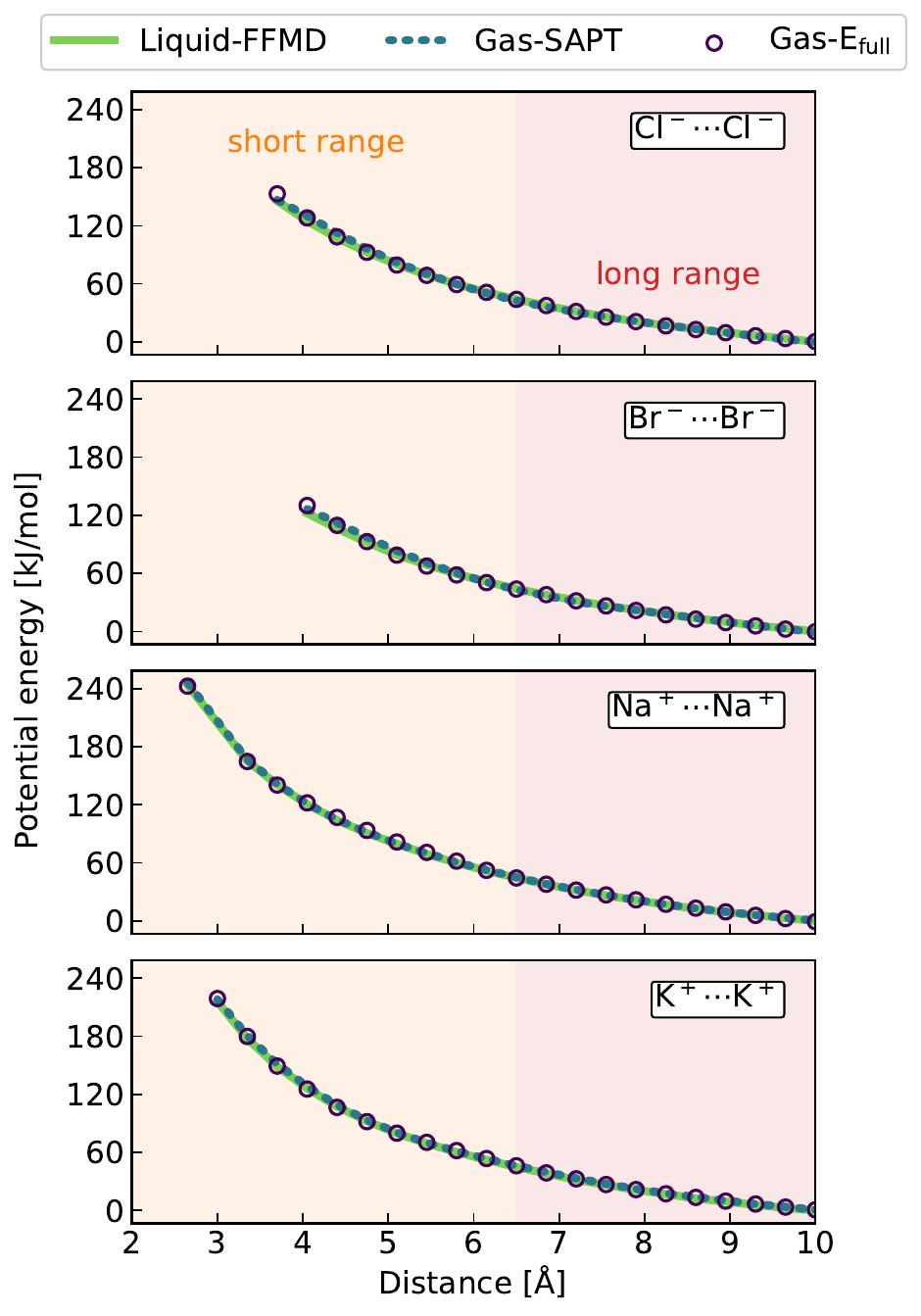}
    \caption{
    Coulombic potential as a function of the interionic separation was obtained by subtracting three distinct FFMD free energies from the AIMD free energy for chloride, bromide, sodium, and potassium from top to bottom.
    Long and short distance range used for the subsequent fitting by the scaled Coulomb potential (Figure~\ref{fig:scaling factors}) is indicated by red and orange background.
    }
    \label{fig:coulombic-potential-extracted}
\end{figure}

The free energies of ion pairing obtained at the FFMD level are also shown in Figure~\ref{fig:free-energies} for the three sets of parameters.
Note that the AIMD and FFMD free energy profiles have different shapes and $y$-scales due to the absence of Coulomb repulsion in the latter, where the ionic charges have been zeroed to extract the non-electrostatic contributions.
The subtraction of the FFMD from the AIMD profiles thus yields the electrostatic contribution (\textit{i.e.}, $U_C$) to the free energy curves which we show in Figure~\ref{fig:coulombic-potential-extracted}.
As a matter of fact, these curves are very similar in shape to the AIMD free energy profiles.
In other words, the interionic Coulombic repulsion dominates the free energy profiles.
Note that the investigated range of interionic separations was cut off at small values corresponding to sizable overlapping electron densities, where strong repulsion between the ions would lead to numerical instabilities.

\begin{figure}[t]
    \centering
    \includegraphics[width=\linewidth]{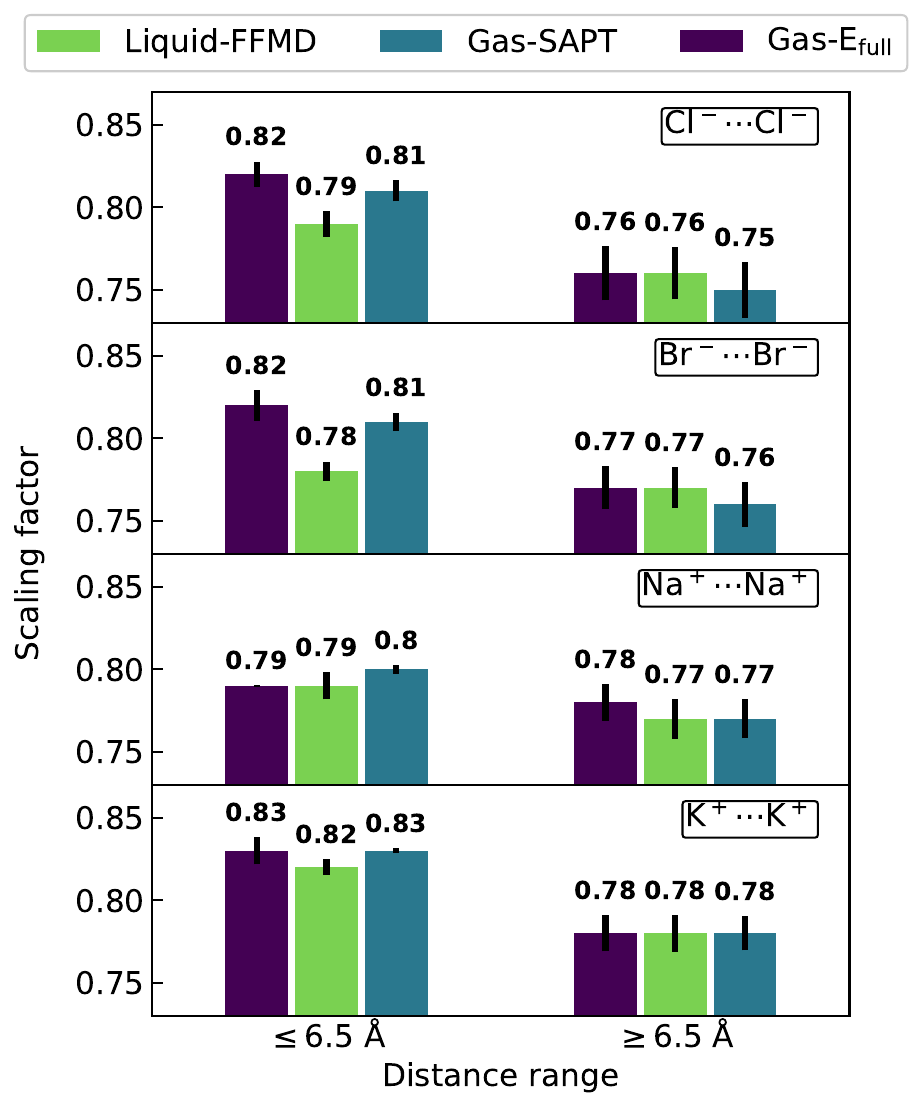}
    \caption
    {
    Scaling factors derived by fitting the extracted Coulombic potentials for shorter and larger ion--ion separations (see SI for the Coulombic curves and the fitting parameters).
    Each color represent one of the three FFMD free energies subtracted from the AIMD free energy yielding the Coulombic potential.
    Error bars indicate standard errors obtained by bootstrapping.
    }
    \label{fig:scaling factors}
\end{figure}

Each of the above-subtracted profiles (Figure~\ref{fig:coulombic-potential-extracted}) was then fitted to a scaled charge Coulomb potential.
More precisely, each profile was first divided into two regions of approximately equal lengths, \textit{i.e.}, at 6.5~\AA\ as indicated by the colored areas in Figure~\ref{fig:coulombic-potential-extracted}.
Next, the scaled Coulomb potential from Equation~\ref{eq:scaled-coulomb} was fitted separately to each of these two regions.
In practice, the potential was first linearized by replotting as a function of the inverse distance, and a line with a slope $a$ and an intercept $b$ was then determined by a least square fit.
Values of $a$ and $b$ and the residuals are provided in Table~S2.
The scaling factor was then extracted from the slope as
\begin{equation}
    s = \sqrt{a\frac{4\pi\varepsilon_0}{q_1q_2}}.
\end{equation}
and plotted in Figure~\ref{fig:scaling factors} for the two regions of the four ion pairs (for the three FFMD free energies used for the subtraction).

For all ion pairs and subtraction schemes, scaling factors~($s$) from 0.76 to 0.78 were obtained for larger ion--ion separations.
This almost constant value provides direct computational evidence that the Coulombic interaction between ions in solution is indeed attenuated as quantitatively described by the ECC approach.
Moreover, the scaling factor at these larger separations is in quantitative agreement with the value following from the experimental dielectric permittivity of the liquid argon solvent.
Note that this value has an experimental uncertainty leading to scaling factors in the range of ($s=[0.75$--$0.81]$)~\cite{CRCHandbook-2016-refraction-index,CRCHandbook-2016-dielectric-constant}.
Another observation is that the scaling factors at closer distances tend to increase slightly compared to those at larger distances.
Namely, short-range scaling factors vary from $s=0.78$ to $s=0.83$.
The main lesson learned from this exercise is that solvent dielectric screening is almost as efficient at smaller interionic separations (including such close proximity that no solvent particles can squeeze between the ions) as at larger ones.
Our computational results thus lend validity to the mean-field ECC approximation employing uniformly scaled charges irrespective of the interionic separation.

In summary, we have employed a combination of AIMD and auxiliary FFMD to quantify the effect screening of ionic charges due to the electronic polarization of the surrounding solvent.
This effective screening manifests itself as a scaled Coulombic interaction, with the scaling factor obtained from simulations being in excellent agreement with the value deduced from the experimental solvent dielectric permittivity, as proposed by the ECC mean-field approach.
The present findings have implications reaching beyond the present model systems of like-charge ion pairs in liquid argon.
The technique of scaling integer charges by the reciprocal square root of the electronic permittivity of the solvent environment lies at the core of the ECC approach aimed at incorporating electronic polarization effects in a mean-field way in complex biological systems.
Furthermore, we show that at short interionic separations, the scaling factor slightly increases up to 10~\%.
Using the scaling factor derived directly from the high-frequency dielectric constant should thus yield good results for both nonpolar and polar systems.
Nevertheless, the above result also lends credibility to using slightly larger scaling factors than those derived from the high-frequency dielectric constant, particularly when the focus is on short interionic separations.
Also, as strong electrostatic screening in polar solvents makes long-range interactions virtually identical for the considered ranges of scaling factors, short-range scaling factors become more relevant in these systems.
Overall, our study provides the ECC framework with a solid theoretical foundation and robust benchmarks, which should aid its further development and broadening of its applicability.

\section*{\label{sec:methods}computational details}

\subsection*{\label{sec:MD-simulations} Common Molecular Dynamics Simulations parameters}

The simulated systems contained a pair of ions of the same type solvated by 512 argon atoms in a cubic unit cell of 28.93~\AA\ side length, determined from the density of neat liquid argon at 1~bar of roughly 35250~mol/m$^3$ (21.228~molecules/nm$^3$)\cite{NIST-argon-2022}, using three-dimensional periodic boundary conditions.
To check for (and exclude the effects of) finite size effects, we also tested systems with 128, 256, and 8192 argon atoms, where cutoffs were adjusted to the resulting box size.
Simulations were carried out in a canonical ensemble at 300 K maintaining the dense supercritical liquid state by not letting the volume expand.
The elevated temperature was employed in order to enhance sampling, being justified by the fact that the polarizability, which is the key target of this study, is a function only of the number density (within the applicability of the Clausius--Mossotti relation).
For the FFMD production simulations, a 9~ns trajectory was acquired after 1~ns equilibration, while for the AIMD production simulations, the system was equilibrated for 5~ps and then propagated for 25~ps per free energy window.

\subsection*{\label{sec:aimd-setup}Ab initio molecular dynamics setup}

\textit{Ab initio} molecular dynamics (AIMD) simulations were carried out with the CP2K software (versions 8.1 and 9.1) using the Quickstep module~\cite{Kuhne2020-10.1063/5.0007045} to employ the hybrid Gaussian and plane waves approach~\cite{Lippert1997-10.1080/002689797170220}.
Atomic nuclei were propagated classically with a 0.5~fs time step, while temperature was controlled by the stochastic velocity rescaling (SVR) thermostat with a time constant of 50~fs during equilibration and 200~fs during production runs.
Electronic structure was calculated at each MD step by the revPBE-D3~\cite{Perdew1996-10.1103/PhysRevLett.77.3865, Zhang1998-10.1103/PhysRevLett.80.890, Goerigk2011-10.1039/C0CP02984J} generalized gradient approximation (GGA) density functional.
The pairwise D3 correction was employed while disabled specifically for all the pairs involving cations~\cite{Kostal2023-10.1021/ACS.JPCLETT.3C00856}.
The orbitals were expanded into the TZV2P Gaussian basis set~\cite{VandeVondele2007-10.1063/1.2770708}, density into plane-wave basis with a 600~Ry energy cutoff, and the core electrons were represented by the Goedecker--Teter--Hutter (GTH) pseudopotentials~\cite{Goedecker1996-10.1103/PhysRevB.54.1703}.
SHAKE/RATTLE algorithm~\cite{Andersen1983-10.1016/0021-99918390014-1} constrained the distance between the ions.

\subsection*{\label{sec:ffmd-setp}Force field molecular dynamics setup}

The force field molecular dynamics (FFMD) simulations were executed with the program GROMACS~2022.2~\cite{Abraham2015-10.1016/J.SOFTX.2015.06.001}.
We used a 2~fs time step during production runs with temperature maintained using the stochastic velocity scaling thermostat~\cite{Bussi2007-10.1063/1.2408420} with a time constant of 1~ps.

A Lennard-Jones (12--6) potential was employed to account for the modeled van der Waals interactions, including all but Coulombic interactions.
The Lennard-Jones parameters for all pairs were obtained by several approaches:
\begin{enumerate}
    \item Liquid-FFMD:  Ion parameters optimized for aqueous systems from Reference~\citenum{Kohagen2014-10.1021/jp5005693} for chloride, \citenum{Prosecco-ions} for bromide, \citenum{Kohagen2016-10.1021/acs.jpcb.5b05221} for sodium and \citenum{Mason2012-10.1021/jp3008267} for potassium.
Additionally, OPLS-AA parameters were used for argon~\cite{Verlet1972-10.1080/00268977200102111}.
Cross terms were obtained using combination rules (arithmetic average for $\sigma$, geometric average for $\varepsilon$).
    \item Gas-SAPT: Interaction energy curves obtained by symmetry-adapted perturbation theory~\cite{Szalewicz2012-10.1002/wcms.86} (SAPT) in the gas phase using def2-TZVPPD~\cite{Rappoport2010-10.1063/1.3484283} basis set and Hartree--Fock wavefunction in Q-Chem~5.3.2.~\cite{Epifanovsky2021-10.1063/5.0055522}.
Note that the electrostatic and polarization contribution to the interaction energy was removed in all cases.
    \item Gas-E$_\mathrm{full}$: Gas-phase full interaction energy curves obtained at the save level of theory as the bulk simulations and wavelet Poisson solver~\cite{Genovese2006-10.1063/1.2335442}.
The interaction energy between species $A$ and $B$ was calculated over a range of distances as
\begin{equation}
    \label{eq:interaction-energy}
    E_\mathrm{AB} = E_\mathrm{AB}^\mathrm{AB} - E_\mathrm{A}^\mathrm{AB} - E_\mathrm{B}^\mathrm{AB},
\end{equation}
where the subscripts denote the system, and the superscript indicates the employed basis set.
Notably, this calculation compensates for basis set superposition errors~\cite{Boys1970-10.1080/00268977000101561}.
For the ion--ion case, a vacuum Coulomb potential was subtracted from the interaction energy scan to obtain only the van der Waals interaction.
\end{enumerate}

Parameters from all three approaches and the corresponding potential energy curves are provided in Section~S4.

\subsection*{\label{sec:free-energy-methods} Free Energy Calculation}

The AIMD free energy profiles of ion pairing were evaluated by the blue moon ensemble approach~\cite{Trzesniak2007-10.1002/CPHC.200600527, Tuckerman2010-978-0-19-852526-4}.
The mean force between two ions was calculated for a series of 19--21 windows of increasing interionic distance, which was constrained in each window.
The mean force between the two ions in each simulation step was computed as an average of the magnitudes of the force vectors of each ion projected onto the displacement line of the two ions.
The free energy profile was then obtained by integrating the mean force along the interionic distance $r$ using the cumulative trapezoidal rule.
For FFMD the free energies were obtained by the accelerated weighted histograms method~\cite{Lundborg2021-10.1063/5.0044352} as implemented in Gromacs~2022.2.
To account for volume entropy, a correction of $2k_BT\ln(r)$ was added to all the free energy profiles, where $k_B$ is the Boltzmann constant and $T$ is the temperature~\cite{Timko2010-10.1063/1.3360310}.

\subsection*{\label{sec:PMF-decomposition} Free Energy Decomposition}

The Coulombic charge--charge interaction potentials between each two ions were extracted from the AIMD free energy profiles using an auxiliary FFMD simulation of the same composition as follows.
First, we write the free energy in the canonical ensemble as
\begin{equation}
    A = U - TS
\end{equation}
where $U$, $T$, and $S$ represent potential energy, temperature, and entropy.
In our case, $U$ can be written as a sum of pairwise Coulomb (C) and van der Waals (vdW) contributions
\begin{equation}
    \label{eq:U-decomposition}
    U = U_\mathrm{C, ion-ion} + \sum_i U_\mathrm{vdW, i}
\end{equation}
where $i$ runs over all atom pairs in our system: \textit{i.e.}, ion--ion, ion--argon and argon--argon.
Note, that this decomposition holds fully for the pairwise empirical force field, but only approximately for the many-body \textit{ab initio} potential.
Next, we simulate the same system both at AIMD and FFMD levels.
For the latter, we use the three pair potential vdW parameters sets (see above), putting zero charges on the ions, causing the first term on the right-hand side of Equation~\ref{eq:U-decomposition} to vanish.
Then we assume that
\begin{equation}
    \sum_i U^\mathrm{AIMD}_\mathrm{vdW, i} \approx \sum_i U^\mathrm{FFMD}_\mathrm{vdW, i}.
\end{equation}
This is an acceptable approximation when considering that the three sets of FFMD parameters were designed to be as compatible as possible with respect to the AIMD calculations.
Moreover, the vdW term is much smaller than the Coulomb one, which further justifies the present approach.
Finally, realizing that the entropy of both systems is essentially the same (\textit{i.e.}, $S^\mathrm{AIMD} \approx S^\mathrm{FFMD}$), we obtain the Coulombic potential contribution to the AIMD free energy as
\begin{equation}
    \label{eq:UC-subtraction}
    U^\mathrm{AIMD}_\mathrm{C, ion-ion} = A^\mathrm{AIMD} - A^\mathrm{FFMD}_\mathrm{zero\ charge}.
\end{equation}

\subsection*{Error estimations}

The error of the mean force from the AIMD was estimated for each window by the bootstrapping method.
The free energy error ($\sigma$) was then obtained by summing the errors of the underlying forces.

Errors of scaling factors were also calculated using the bootstrapping method.
For evaluation of the AIMD forces, trajectories were divided into 1~ps intervals, and resampling was performed 1000 times.
In each cycle, the Coulombic potential was extracted, and the scaled Coulomb law was used to fit and obtain the scaling factors.

The standard error reported throughout this paper was calculated as $1.96\sigma$, corresponding to a 95~\% confidence interval.

\section*{Supporting Information}
Extracted Coulombic potential curves together with their fitted scaling factors and parameters of the fitting, evaluation of the finite box size effect on the free energies, error estimation of the scaling factors, and pair potential energy curves obtained by the three different methods with their Lennard-Jones parameters.

\begin{acknowledgements}

We thank Jan Řezáč for fruitful discussions regarding the SAPT method.
V.K. acknowledges Faculty of Science of Charles University where he is enrolled as a Ph.D. student and the IMPRS for Quantum Dynamics and Control in Dresden.
This work was supported by the Ministry of Education, Youth and Sports of the Czech Republic through the e-INFRA CZ (ID:90140). P.J. acknowledges support from the European Research Council via an ERC Advanced Grant no. 101095957.

\end{acknowledgements}

\section*{References}

%aipnum4-2.bst 2019-01-14 (MD) hand-edited version of apsrev4-1.bst

\end{document}

\def\mstitle{Supporting Information for: Non-Aqueous Ion Pairing Exemplifies the Case for Including Electronic Polarization in Molecular Dynamics Simulations}
\title{\mstitle}

\author{Vojtech Kostal}
\affiliation{
Institute of Organic Chemistry and Biochemistry of the Czech Academy of Sciences, Flemingovo nám. 2, 166 10 Prague 6, Czech Republic
}

\author{Pavel Jungwirth*}
\email{pavel.jungwirth@uochb.cas.cz}
\affiliation{
Institute of Organic Chemistry and Biochemistry of the Czech Academy of Sciences, Flemingovo nám. 2, 166 10 Prague 6, Czech Republic
}

\author{Hector Martinez-Seara*}
\email{hseara@gmail.com}
\affiliation{
Institute of Organic Chemistry and Biochemistry of the Czech Academy of Sciences, Flemingovo nám. 2, 166 10 Prague 6, Czech Republic
}

\date{\today}

\maketitle

\section{Finite size effects}

In this study, the systems under investigation are simulated with a net charge, thus opposite-sign charges are distributed on the walls of the simulation box to handle long-range electrostatic interactions.
Therefore, an error is introduced when calculating free energies.

To quantify this error, we conducted calculations of the free energy of ion pairing (FFMD) as a function of distance in boxes of different sizes ranging from 128 to 8192 solvent molecules.
Specifically, we utilized a pair of chloride anions with the same Lennard-Jones parameters as mentioned in the main text, and a charge of $-0.75$. Cutoffs where adjusted to the resulting box size.

\begin{figure}[b]
    \centering
    \includegraphics[width=\linewidth]{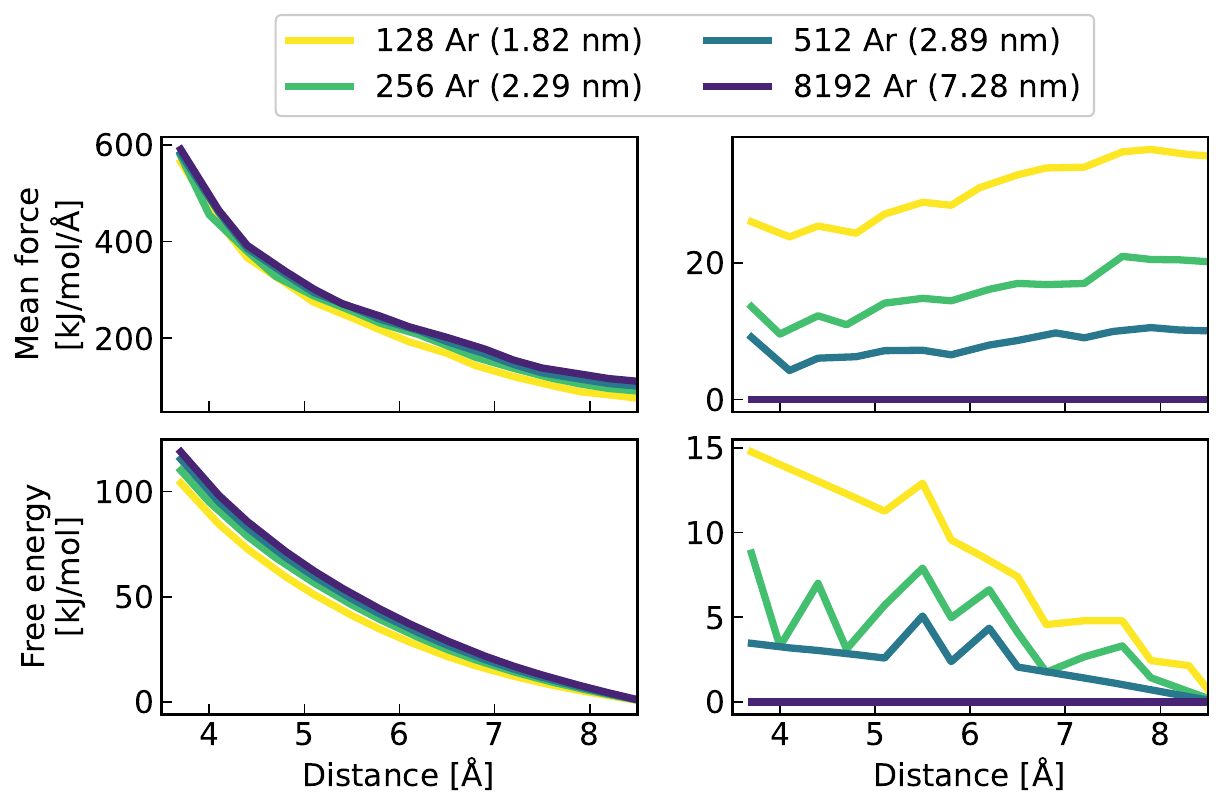}
    \caption
    {
    Mean forces (top left) and free energies (bottom left) of the Cl--Cl pairing as a function of distance for systems of varying sizes.
    The differences between the results for the largest system and the smaller ones are displayed in the panels on the right-hand side. The box sizes are indicated in parentheses.
    }
    \label{fig:si-pme-error}
\end{figure}

Figure~\ref{fig:si-pme-error} demonstrates a noticeable difference in mean forces and free energy as the system size increases.
However, for the system consisting of two chloride anions and 512 solvent molecules, which is the case throughout this study, the error in free energy is below 5 kJ/mol (less than 5\% of the total free energy.
Based on this observation, we conclude that a system containing 512 solvent molecules is sufficiently large to mostly mitigate the effects of finite-size boxes.

\section{Errors of the scaling factors}

Table~\ref{tab:scaling-errors} shows standard errors of the scaling factors from Figure~1 calculated as described in Computational Details in the main text.

\begin{table}[htbp]
    \centering
    \caption{Errors of the scaling factors from Figure~1 of the main text.}
    \label{tab:scaling-errors}
    \begin{tabular}{rcccc}
        \toprule
        & Region & MD & SAPT & E$_\mathrm{int}$ \\
        \midrule
        \multirow{2}{*}{Cl$\cdots$Cl} & 1 & 0.0158 & 0.017 & 0.0164 \\
         & 2 & 0.0079 & 0.0064 & 0.0077 \\
         \midrule
        \multirow{2}{*}{Br$\cdots$Br} & 1 & 0.0124 & 0.0136 & 0.013 \\
         & 2 & 0.0059 & 0.0056 & 0.0094 \\
         \midrule
        \multirow{2}{*}{Na$\cdots$Na} & 1 & 0.012 & 0.0117 & 0.0113 \\
         & 2 & 0.0083 & 0.0027 & 0.0006 \\
         \midrule
        \multirow{2}{*}{K$\cdots$K} & 1 & 0.0111 & 0.0104 & 0.0109 \\
         & 2 & 0.0048 & 0.0015 & 0.0082 \\
    \end{tabular}
\end{table}

\section{Fitting the Coulombic interaction}

\begin{figure*}[b]
    \centering
    \includegraphics[width=\linewidth]{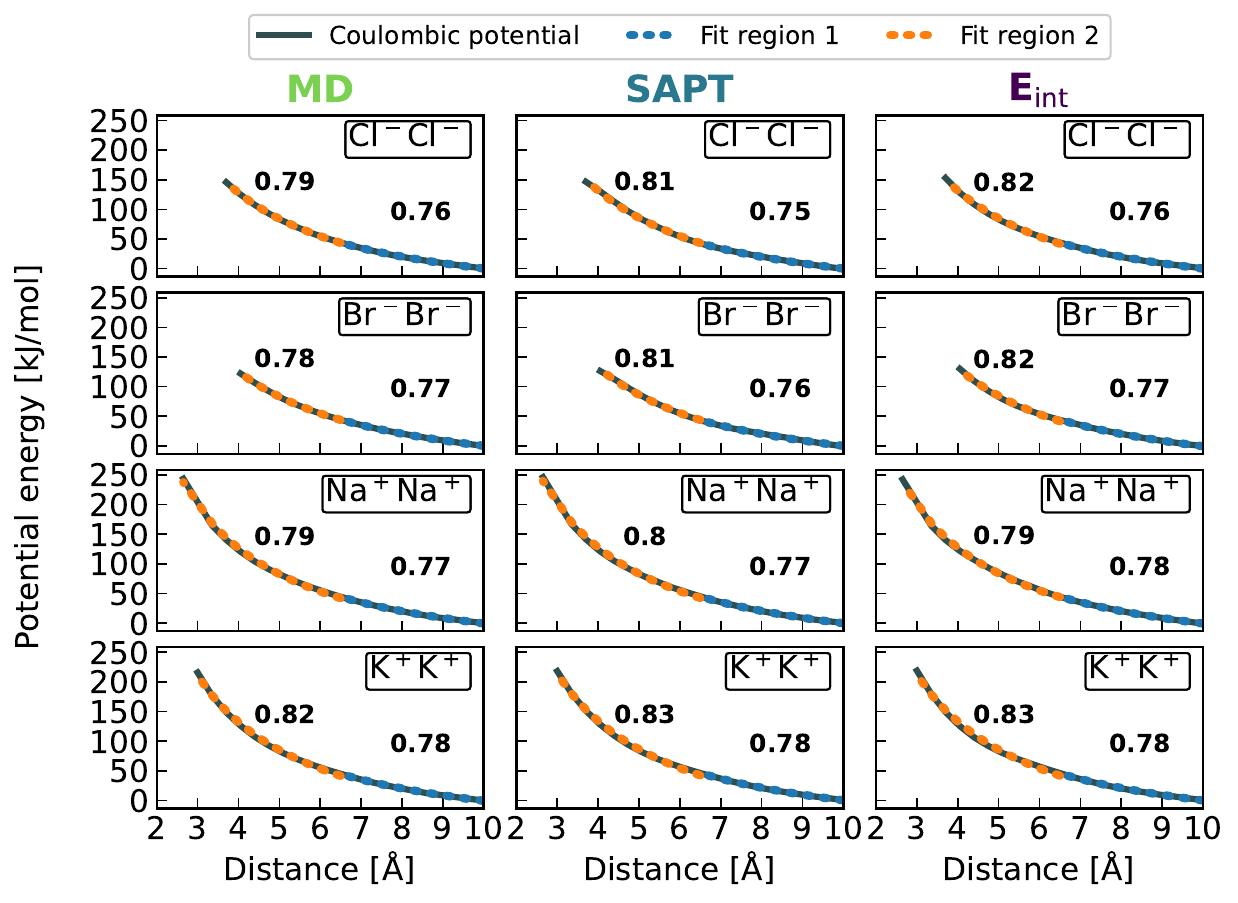}
    \caption{
        Coulombic potential obtained as a difference between AIMD and FFMD free energy profile (full grey line) for different ions (columns) and various FFMD free energies (rows).
        Fitted scaled Coulomb potential is plotted as a dashed line for the larger distances region (1, blue) and for the short distances region (2, orange). 
    }
    \label{fig:si-coulomb-fit}
\end{figure*}

To compare the AIMD results with the theory-predicted scaled Coulomb law, the Coulombic interaction was initially extracted from the AIMD free energy by subtracting the corresponding FFMD free energy as described in the Computational Details of the main text.
The subtracted curves are shown in Figure~\ref{fig:si-coulomb-fit}.

Then, the extracted curve was linearized by plotting it against the reciprocal distance for two distinct regions covering below and above 6.5~\AA.
Each region was then fitted by a line.
Scaling factors were then obtained from the slope of the linear fit and are shown above each region.
Parameters of the linear fit for each system and region are provided in Table~\ref{tab:linear-fit-params}.

\begin{table*}[b]
\centering
\label{tab:linear-fit-params}
\caption{Slopes ($a$), intercepts ($b$), and residuals ($R^2$) values for all the linear fits of the scaled Coulomb law (Equation~1) to the extracted Coulombic potentials.}
\begin{tabular}{lcccccccccc}
    \toprule
    & & \multicolumn{3}{c}{MD} & \multicolumn{3}{c}{SAPT} & \multicolumn{3}{c}{E$_\mathbf{int}$} \\
    \cmidrule(lr){3-5}
    \cmidrule(lr){6-8}
    \cmidrule(lr){9-11}
    & Region & $a$ & $b$ & $R^2$ & $a$ & $b$ & $R^2$ & $a$ & $b$ & $R^2$ \\
    \midrule
    \multirow{2}{*}{Cl$\cdots$Cl} &
       1 & $80.4232$ & $-80.2171$ & $0.9995$ & $87.4153$ & $-90.9885$ & $0.9999$ & $80.6331$ & $-80.5952$ & $0.9995$ \\
     &  2 & $78.9750$ & $-78.8471$ & $0.9999$ & $91.6763$ & $-97.8888$ & $0.9982$ & $92.9286$ & $-101.0818$ & $0.9978$ \\
    \midrule
    \multirow{2}{*}{Br$\cdots$Br} &
       1 & $81.6126$ & $-81.4234$ & $0.9999$ & $84.4547$ & $-86.0647$ & $0.9999$ & $82.7884$ & $-83.2828$ & $0.9998$ \\
     &  2 & $80.0871$ & $-79.9229$ & $0.9982$ & $91.5796$ & $-97.1833$ & $0.9981$ & $92.7008$ & $-101.0489$ & $0.9977$ \\
    \midrule
    \multirow{2}{*}{Na$\cdots$Na} &
       1 & $82.7369$ & $-83.1086$ & $0.9996$ & $87.5174$ & $-92.4609$ & $0.9996$ & $83.2219$ & $-83.3397$ & $0.9995$ \\
     &  2 & $83.2219$ & $-83.3397$ & $0.9996$ & $88.3134$ & $-93.3096$ & $0.9996$ & $86.4249$ & $-88.9900$ & $0.9999$ \\
    \midrule
    \multirow{2}{*}{K$\cdots$K} &
       1 & $83.6421$ & $-84.2042$ & $0.9994$ & $94.1107$ & $-103.2969$ & $0.9994$ & $83.9786$ & $-83.9313$ & $0.9990$ \\
     &  2 & $84.7755$ & $-84.4773$ & $0.9991$ & $95.2801$ & $-104.2518$ & $0.9991$ & $94.7895$ & $-83.9864$ & $0.9998$ \\
  \end{tabular}
\end{table*}

\newpage
\section{Gas-phase pair interaction potentials}

\begin{figure*}
    \includegraphics[width=0.8\linewidth]{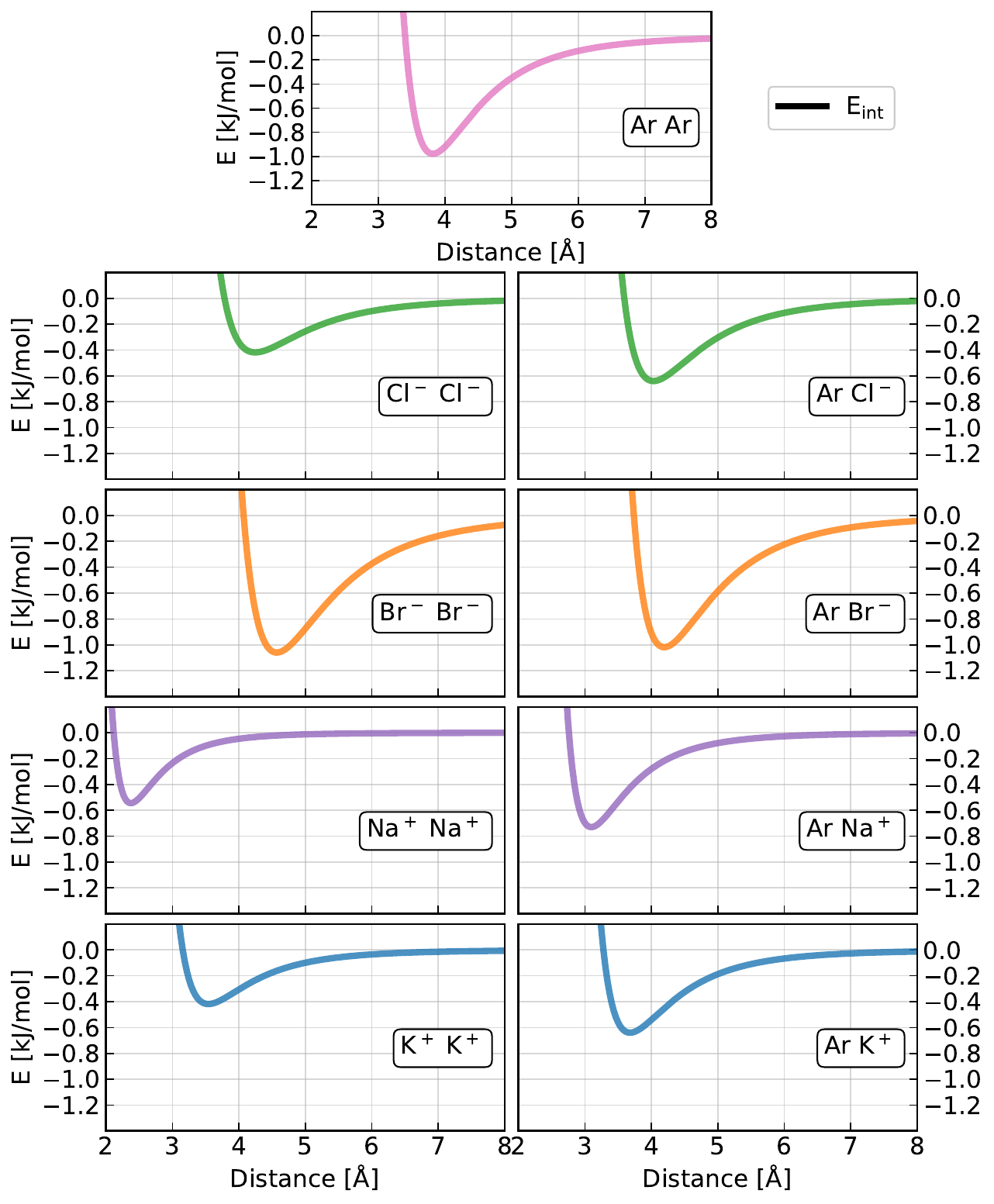}
    \caption{Lennard-Jones (12--6) potential energy curves for all pairs that occur in the presented study.
    Employed parameters are listed in Table~\ref{tab:nb-params} under the MD section.
    Note that the pair potentials were obtained using arithmetic, geometric combination rule for sigma, epsilon respectively.}
    \label{fig:eint-md}
\end{figure*}

\begin{figure*}
    \includegraphics[width=0.8\linewidth]{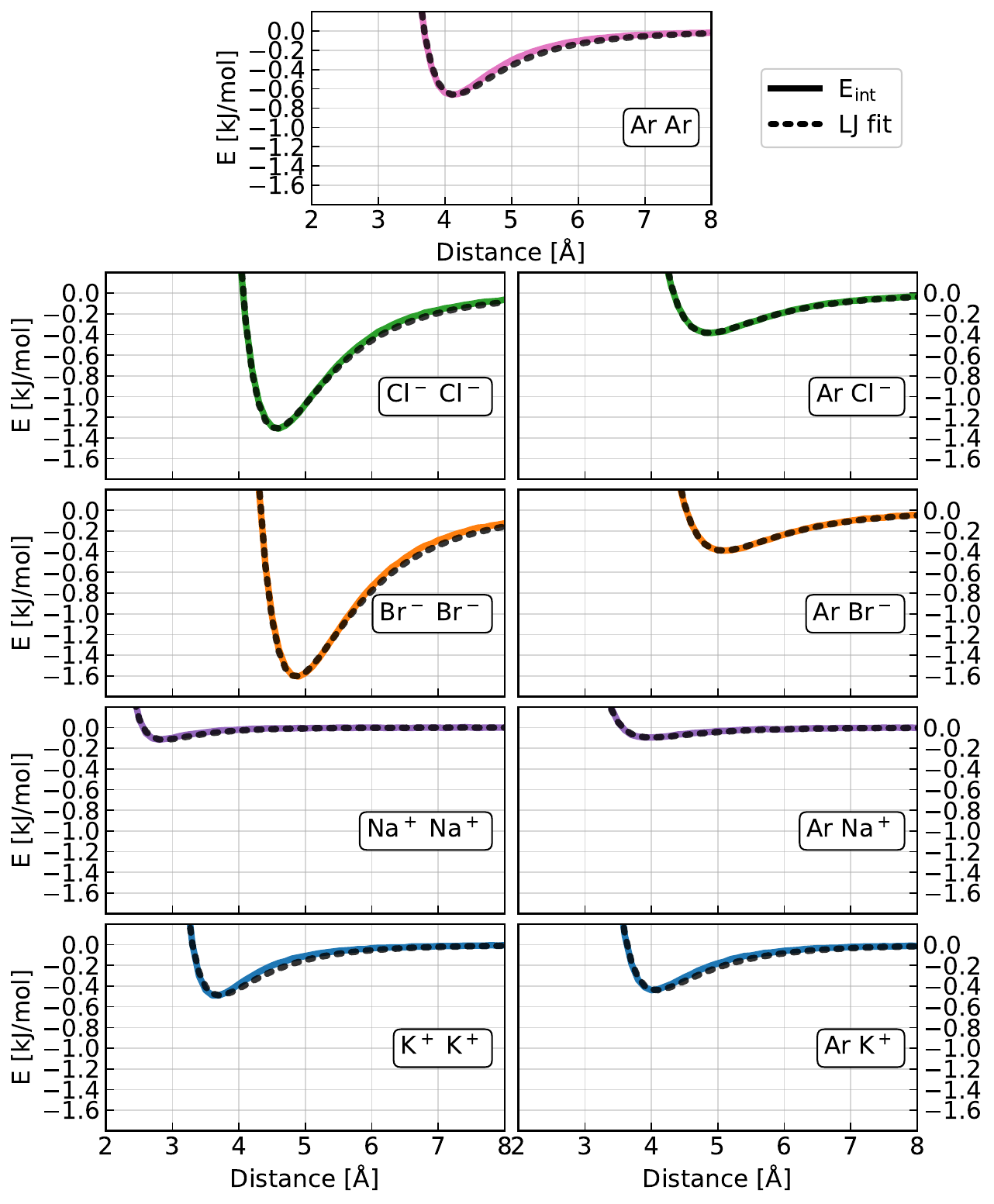}
    \caption{
    Interaction energy curves for all the pairs that occur in the presented study obtained at SAPT level (full lines) and their corresponding Lennard-Jones (12--6) fits as dashed black lines.
    Parameters are listed in Table~\ref{tab:nb-params} under the SAPT section.}
    \label{fig:eint-sapt}
\end{figure*}

\begin{figure*}
    \includegraphics[width=0.8\linewidth]{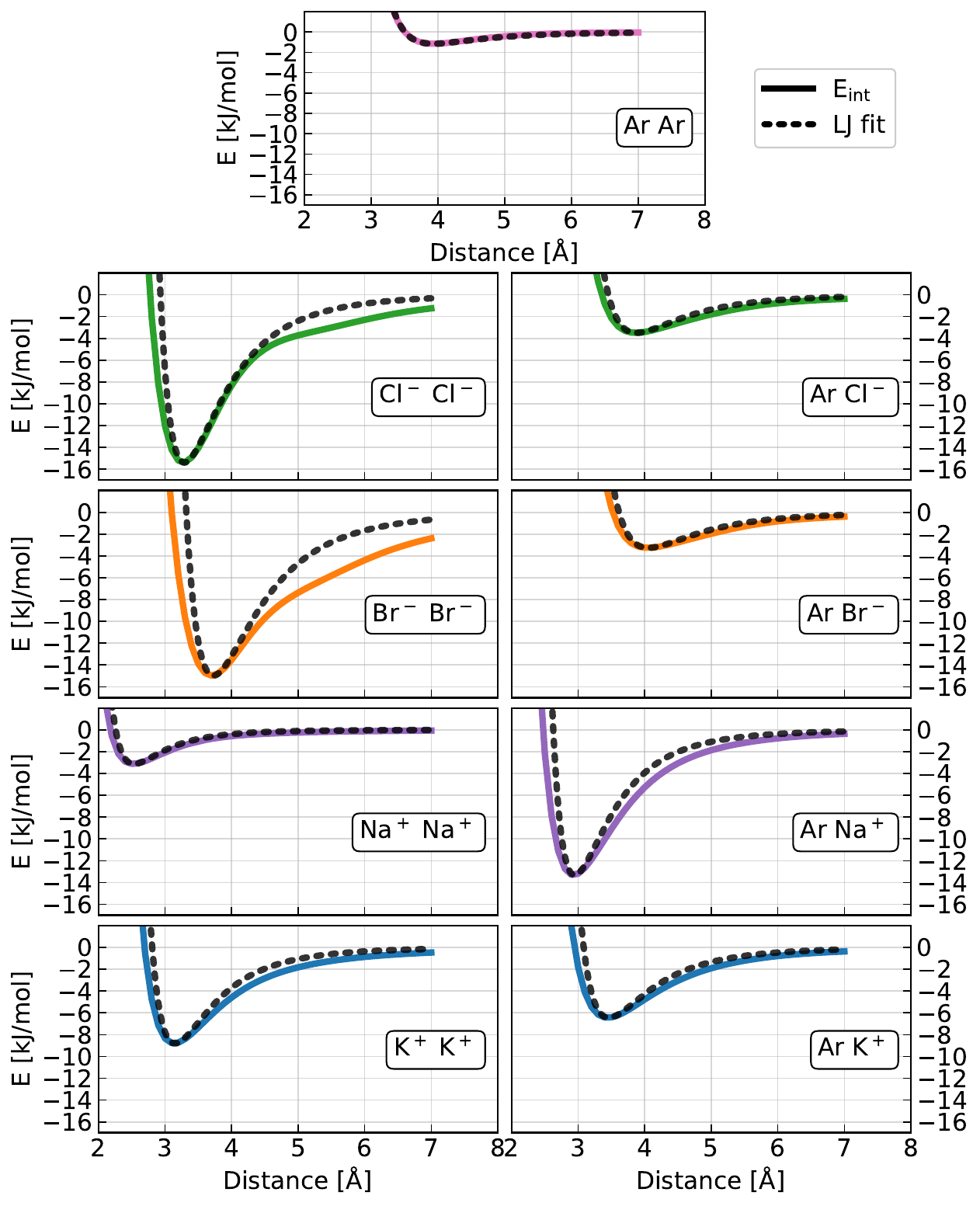}
    \caption{Interaction energy curves for all pairs that occur in the presented study obtained at the revPBE-D3 (Ar$\cdots$Ar, anion$\cdots$Ar, anion$\cdots$anion) and revPBE (Ar$\cdots$cation, cation$\cdots$cation) level (full lines) and their corresponding Lennard-Jones (12--6) fits as dashed black lines.
    Parameters are listed in Table~\ref{tab:nb-params} under E$_\mathrm{int}$ section.}
    \label{fig:eint-cp2k}
\end{figure*}

Figures~\ref{fig:eint-md}, \ref{fig:eint-sapt}, and \ref{fig:eint-cp2k} provide the pair interaction energy curves for all the atomic pairs used in this work together with the fitted Lennard-Jones (12--6) potentials when needed.
Interaction potentials from the Liquid-MD parameters are shown in Figure~\ref{fig:eint-md}, from Gas-SAPT parameters in Figure~\ref{fig:eint-sapt} and from Gas-E$_\mathrm{full}$ in Figure~\ref{fig:eint-cp2k}.
The used or fitted Lennard-Jones parameters are listed in Table~\ref{tab:nb-params}

\begin{table}[h!]
    \centering
    \caption{Lennard-Jones 12--6 potential parameters for all the atom pairs involved in this paper.}
    \begin{tabular}{lrrrrrr}
        \toprule
         & \multicolumn{2}{c}{MD} & \multicolumn{2}{c}{SAPT} & \multicolumn{2}{c}{E$_\mathrm{int}$}\\
         \cmidrule(lr){2-3}
         \cmidrule(lr){4-5}
         \cmidrule(lr){6-7}
         & $\sigma$ [nm] & $\varepsilon$ [kJ/mol] & $\sigma$ [nm] & $\varepsilon$ [kJ/mol] & $\sigma$ [nm] & $\varepsilon$ [kJ/mol] \\
         \midrule
         Ar$\cdots$Ar & 0.3401 & 0.9786 & 0.3689 & 0.6611 & 0.3504 & 1.1530 \\
         Cl$\cdots$Cl & 0.3782 & 0.4184 & 0.4063 & 1.3096 & 0.2931 & 15.3915 \\
         Cl$\cdots$Ar & 0.3592 & 0.6399 & 0.4333 & 0.3849 & 0.3454 & 3.4959 \\
         Br$\cdots$Br & 0.4070 & 1.0600 & 0.4328 & 1.6067 & 0.3319 & 14.9781 \\
         Br$\cdots$Ar & 0.3736 & 1.0185 & 0.4526 & 0.3891 & 0.3614 & 3.2577 \\
         Na$\cdots$Na & 0.2115 & 0.5443 & 0.2559 & 0.1172 & 0.2259 & 3.1199 \\
         Na$\cdots$Ar & 0.2758 & 0.7298 & 0.3567 & 0.0962 & 0.2626 & 13.3235 \\
         K$\cdots$K   & 0.3154 & 0.4187 & 0.3301 & 0.4895 & 0.2807 & 8.8005 \\
         K$\cdots$Ar  & 0.3278 & 0.6401 & 0.3632 & 0.4393 & 0.3083 & 6.4452 \\
    \end{tabular}
    \label{tab:nb-params}
\end{table}

\newpage

\clearpage
\bibliography{bibliography}